\documentclass[nolineno]{aa}
\usepackage{natbib,twoopt}
\usepackage[breaklinks=true]{hyperref} 
\bibpunct{(}{)}{;}{a}{}{,}             

\makeatletter
\newcommandtwoopt{\citeads}[3][][]{\href{http://adsabs.harvard.edu/abs/#3}%
	{\def\hyper@linkstart##1##2{}%
		\let\hyper@linkend\@empty\citealp[#1][#2]{#3}}}
\newcommandtwoopt{\citepads}[3][][]{\href{http://adsabs.harvard.edu/abs/#3}%
	{\def\hyper@linkstart##1##2{}%
		\let\hyper@linkend\@empty\citep[#1][#2]{#3}}}
\newcommandtwoopt{\citetads}[3][][]{\href{http://adsabs.harvard.edu/abs/#3}%
	{\def\hyper@linkstart##1##2{}%
		\let\hyper@linkend\@empty\citet[#1][#2]{#3}}}
\newcommandtwoopt{\citeyearads}[3][][]%
{\href{http://adsabs.harvard.edu/abs/#3}
	{\def\hyper@linkstart##1##2{}%
		\let\hyper@linkend\@empty\citeyear[#1][#2]{#3}}}
\makeatother
\usepackage{graphicx}
\usepackage[varg]{txfonts}
\nolinenumbers

\begin{document}

	\title{Variation of sunspot groups' polarity separation during their evolution}
	
	\subtitle{}
	
	\author{Judit Muraközy
		\inst{1}
	}
	
	\institute{HUN-REN Institute of Earth Physics and Space Science,
		Csatkai Endre str. 6-8, H-9400 Sopron\\
		\email{murakozy.judit@epss.hun-ren.hu}
	}
	
	\date{Received \today; accepted XXX X, X}

\abstract{
During the emergence of sunspot groups the footpoints of their leading and following parts move apart. This diverging motion results in the stretching of the active regions, which continues during the decay phase. The aim of the present work is to study the separation distance variations during the active region evolution on a large statistical sample. Altogether more than 2000 individual sunspot groups were taken into account. The investigation is mainly based on data of the SoHO/MDI - Debrecen Sunspot Data (SDD) catalog which covers the time span 1996-2010, practically the whole solar cycle 23. For check of the possible cyclical variation the Debrecen Photoheliographic Data (DPD) is used which contains data for solar cycles 20-24. The separation distance is calculated between the leading and following centers of mass and starts to increase after the emergence and shows a plateau around the peak flux. The polarity separation reaches its maximum in the decay phase and then starts to decrease in the cases of the largest and medium size groups, but continues its increase in the case of the smallest groups. This decrease is caused by the eastward motion of the leading part, while the following part continues its backward motion. The separation distance is size dependent, i. e., the larger the sunspot group the greater its extent. Cycle and cycle phase dependencies as well as hemispheric connection can also be observed.}

\maketitle
\titlerunning{<short title>}
\authorrunning{<name(s) of
	author(s)}
\section{Introduction}
Studies of the solar active region (AR) evolution are important regarding to the solar dynamo theories. The development and decay of the active region are governed by different mechanisms. While the buoyancy is responsible for the emergence the decay is governed by the small-scale cancellation process which fragments and scatters the magnetic flux \citep{2015LRSP...12....1V}. \citet{2021A&A...647A.146S} analyzed statistically the formation and decay of ARs and found that the decay is governed by an erosion process. Thus their detailed study is important to gain more knowledge about the whole process. 
The emergence is a more known phenomenon in the sense of motion as well. Several works studied the leading--following polarity separation of ARs. \citet{1986ApJ...303..480G} showed that the leading following sunspots can be characterized as a simple expansion along their major axis. Although \citet{2003ApJ...593.1217P} studied only six ARs, the image of the polarity separation continued to be refined. They obtained that the footpoint separation distances increase over time after the AR's emergence and reach a plateau 1.5 days after the beginning of the emergence. \citet{2019A&A...625A..53S} studied the motion of the opposite magnetic polarity regions of sunspot groups during their emergence. They pointed out that two phases of emergence can be distinguished. During the first phase, the polarity separation is accelerating, while during the second one is decelerating. The most striking feature is that the leading part moves towards the equator and westward while the following one moves in the  poleward and eastward direction. \citet{2016ApJ...818....7M} studied the variations of separation distances in sunspot groups of different sizes during their evolution and they found that the expansion behavior was different in sunspot groups of different sizes. \citet{2008ApJ...688L.115K} studied 715 bipolar ARs from SOHO/MDI magnetograms and found that after the emergence of the ARs the separation distance increases rapidly but during the decay it decreases. The process pointed out by \citet{2016A&A...596A...3V} is similar. They investigated the evolution of AR NOAA 12118 and pointed out its four phases: the initial separation of the two polarities followed by a rapid expansion phase, then the spread stalls, and finally the period when the AR slowly shrinks. The separation rate is the highest in the initial stage with a value of 0.26 km s$^{-1}$ and after the halt of the separation decreased.
The decay, i.e., the last phase of the AR's lifetime was also investigated several times. This phase takes longer than the emerging phase \citep{2011SoPh..270..463J} and differs for the leading and following parts. Its rate was calculated by different studies. A part of them reported decay as a linear process \citep{1963BAICz..14...91B, 2014SoPh..289.1531G, 2020ApJ...892..107M, 2021ApJ...913..147L, 2021ApJ...908..133M}, while \citet{2003A&ARv..11..153S} noted that linear decay law is appropriate for 95\% of all of the sunspots regardless whether the sunspot groups are recurrent or not. Other studies refer to decay as a parabolic process \citep{1988A&A...205..289M, 1993A&A...274..521M, 1997SoPh..176..249P, 2015ApJ...800..130L}, where the temporal variation of the area is proportional to the square root of the area.
Investigating sunspot groups of Cycle 23 \citet{2008AdSpR..41..881Z} pointed out that the variations of polarity separation with latitude are in anti-phase with those of the tilts reaching a maximum at the latitude of 35$^{\circ}$.
Although, there are many studies on the leading - following separation distance but mainly for the development phase of the ARs life, and they are sparse during the decay. This study aims the statistical investigation of the polarity separation of sunspot groups, as well as its possible hemispheric, cycle and cycle phase dependencies on a large sample by using the Debrecen sunspot databases. 

\section{Data and method}
The unique Debrecen sunspot databases \footnote{\url{http://fenyi.solarobs.epss.hun-ren.hu/en/databases/Summary/}} \citep{2016SoPh..291.3081B, 2017MNRAS.465.1259G} allow us to study the evolution of sunspot groups in high temporal and spatial resolution. 
Thus, the growth and decay of sunspot groups and their leading and following parts can be tracked separately. This study uses two sunspot catalogs. One of them is the SOHO-MDI/Debrecen Sunspot Data (hereafter SDD) that contains data for the period 1996-2010, i.e., solar cycle 23. The input of this database comes from the Solar and Heliospheric Observatory spacecraft and its Michelson Doppler Imager as a white light image and a magnetogram. The SDD’s temporal resolution is 1.5 hour. Since the SDD contains magnetic data for each sunspot the two opposite polarity parts of the groups can be distinguished. Thus, by using the area, and the latitudinal and longitudinal data of sunspots the centers of mass of the leading and following polarities can be calculated. To take the area distributions of those parts into account the area-weighted latitude and longitude are calculated as 
\begin{equation}
B_x=\frac{\sum{b_x a_x}}{\sum a_x} \hspace{5mm} and \hspace{5mm}
L_x=\frac{\sum{l_x a_x}}{\sum a_x},
\label{eq:posdata}
\end{equation}
where $x$ means the polarity, $a$ denotes the area in millionth of solar hemisphere (MSH) corrected for foreshortening, while $b$ and $l$ indicate the latitude and longitude of the spots, respectively. The leading-following magnetic flux contents, and the sum of their absolute values (f$_U$=|f$_{lU}$|+|f$_{fU}$|) called total unsigned flux are determined in Weber until otherwise stated. The flux is calculated for each sunspots and each observation time. Then the fluxes absolute values are summarized for the whole sunspot group and separately its leading and following parts. To check the possible cycle dependence of the polarity separation the tilt angle database \citep{2015MNRAS.447.1857B} of the so-called DPD (Debrecen Photoheliographic Data) sunspot catalog has been used. Its time span covers the period 1974 – 2018, solar cycles 20 - 24. The data of this catalog based on ground-based observation, thus there is no magnetic data in it and its temporal resolution is 1 day. For the present study the DPD tilt angle database is used because this contains the values of the leading and following parts’ centers of mass. In the lack of magnetic data the position of the leading/following part is considered as the area-weighted position of those sunspots that precede/lag the center of mass of the whole sunspot group, respectively. 

The polarity separation has been calculated as the spherical distance between the leading and following centers of mass
\begin{equation}
\delta_U=acos(sin B_l sin B_f + cos B_l cos B_f cos (L_f-L_l))
\label{eq:sphdist}
\end{equation}
where B and L mean the area-weighted latitude and longitude of the leading and following parts (l and f lower indices), respectively. The separation distance is measured in degrees.
Since the sunspot groups with different area evolve in different way and at different speed their areas must be eliminated to compare them. Therefore instead of time, the areal evolution phase \citep{2022ApJ...925...87M} will be used
\begin{equation}
AEP=\left(1-\frac{a}{A}\right)100
\label{eq:adp}
\end{equation}
where $a$ is area of each sunspot group while $A$ is their peak area. In order to the growth and decay phases can be distinguished the value of the AEP and FEP are multiplied by -1 for the growth. Thus, its value is negative during the growth phase of the group, positive if the group decays, and 0 at the time of the maximum area. The flux evolution phase (FEP) can be calculated similarly. These are determined for each observational time of the groups by taking into account the umbral area or the flux. In order to extend the previously used statistical sample (detailed in \citet{2022ApJ...925...87M}) those sunspot groups are considered which show a clear growth phase before and decay phase after the peak area at least 1 day. The number of sample selected such a way from the SDD is 1014, while altogether 1047 groups were studied from the DPD. The evolutionary process is tracked when the groups are within 0.75 of solar radius. Each active region is considered during one transit on the solar disk regardless whether it is recurrent or not. In the present paper only umbrae are taken into account as denoted by the subscript U.

\section{Results and Discussion}
In the previous paper \citep{2022ApJ...925...87M} the decay phase has been studied on 142 selected and visually inspected sunspot groups. Although I also carried out the tests detailed below on the previous sample, only the results based on the extended sample will be shown here. Otherwise, the results are similar for both data sets. 
\subsection{Magnetic Flux}
\begin{figure}[h]
	\centering
	\includegraphics[width=\hsize]{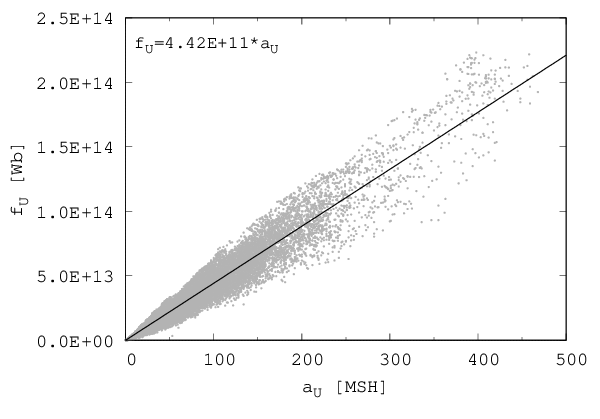}
	\caption{Total unsigned magnetic flux measured in Weber as a function of the area measured in Millionths of Solar Hemisphere. The figure contains the data of all studied sunspot groups from SDD, each gray point marks a sunspot group. The linear fitted function is depicted with solid black line. Equation of the fitted linear function can be seen in the top left corner.}
	\label{Fig:fa}
\end{figure}

The evolution of an active region can be described by the variation of either the total umbral area or the total unsigned magnetic flux measured in the umbrae, at first their relationship is studied. Here the total unsigned magnetic flux means the whole flux that is contained by the leading and following sunspots (f$_U$=|f$_{lU}$|+|f$_{fU}$|) disregarding their polarity. Figure~\ref{Fig:fa} shows the total unsigned ubmral magnetic flux measured in Wb as a function of umbral area.

\noindent
As the plot shows their relationship is linear; the higher the area, the higher its flux content. The equation of the fitted linear function can be seen in the figure.
This relationship is previously investigated by \citet{1966ApJ...144..723S} by considering 8 sunspot groups and obtained that $\phi$$=1.2A$ where $A$ is the area measured in 10$^{18}$cm$^2$, and $\phi$ is expressed in 10$^{21}$Mx. The result of the present study is $\phi=1.47A_U$ if $A_U$ and $\phi$ are expressed in the mentioned units and in good agreement with \citet{1966ApJ...144..723S}. 

\begin{figure}[h]
	\centering
	\includegraphics[width=\hsize]{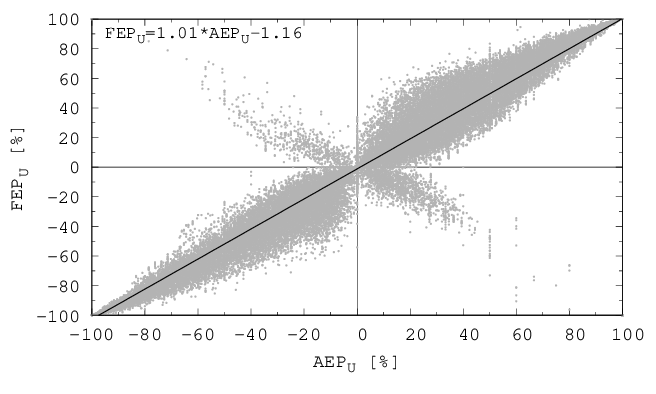}
	\includegraphics[width=\hsize]{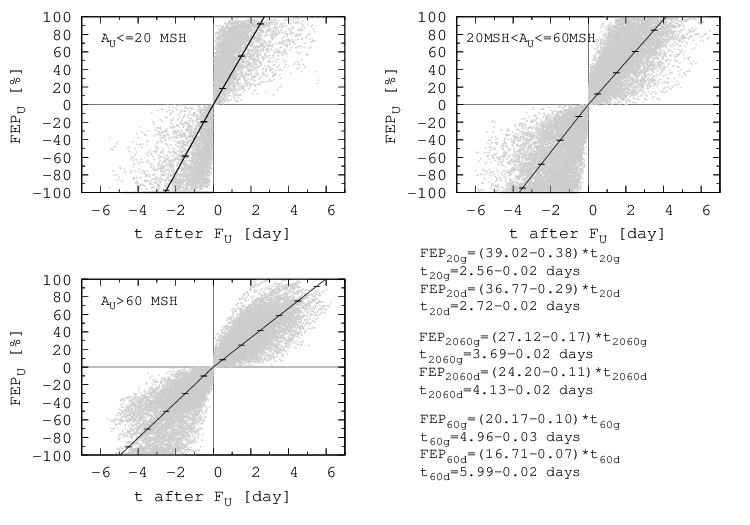}
	\caption{Relationship of the area evolution phase and total unsigned flux evolution phase based on the SDD (top panel). Both of them are measured in percentage. Each gray point marks a sunspot group in one observation time. The linear fitted function is depicted with solid black line, its equation can be found in the upper left quadrant. Bottom panel: the umbral flux evolution phase as a function of time near the time of maximum flux measured in days for three sets of sunspot group sizes. The equations of the fitted regression lines and standard error values can be found in the bottom right quarter. The characters g and d mean the growth (negative values) and decay (positive values) phases of groups. The average growth and decay periods are calculated by using the equations of the fitted functions.}
	\label{Fig:adpfdp}
\end{figure}

The top panel of Figure~\ref{Fig:adpfdp} depicts the AEP - FEP relationship. The bulk of the sunspot groups are mostly located in two quadrants, i.e. positive FEP and AEP or negative FEP and AEP. This means that sunspot groups are in their same phases (growth or decay) at the same time considering the area and flux. Some groups have different times of peak area and peak flux. The number of these cases is 379 while that of the coinciding peak times is 635 in the selected sample. The fitted linear function has an intercept on the vertical axis at -1.16 meaning that the majority of groups reach their peak area slightly earlier than their maximum flux.
As the DPD does not contain magnetic polarity data the AEP will be used for comparison. In the bottom panel of Figure~\ref{Fig:adpfdp} can be seen the connection between the FEP$_U$ and the time near the time of maximum flux.  Since the sunspot groups with different peak area grow and decay in different ways \citep{2014SoPh..289..563M, 2020ApJ...892..107M, 2021ApJ...908..133M}, here we consider three sets of area ranges of sunspot groups. Their peak sizes are smaller than 20 MSH, between 20 MSH and 60 MSH, and larger than 60 MSH (indicated in the panels). Each gray point marks a sunspot  group in one observation time. The negative and positive values mean the growth and decay phases, respectively, the flux maximum is at the zero value of the time axis. Linear functions have been fitted to the growth and decay phases separately. Their equations are in the bottom right quadrant and the growth and decay periods calculated from those equations. Taking into account the peak area of sunspot groups one can conclude that the mean period of the growth phase is shorter than the decay phase in each case. The growth phases are 2.56$\pm$0.02 days, 3.69$\pm$0.02 days and 4.96$\pm$0.03 days for the smallest, middle size and the largest groups, respectively. While the decay phases are 2.72$\pm$0.02 days, 4.13$\pm$0.02 days and 5.99$\pm$0.02 days for the smallest, middle size and the largest groups, respectively. These are different representations of the asymmetric shape of the active region evolution which are referred to by several authors, e.g., \citet{2011SoPh..270..463J}.

\subsection{Polarity Separation}
The separation distance is calculated between the area weighted positions of the centers of mass of sunspots with leading and following polarities by using Equation~\ref{eq:sphdist}.
 
\begin{figure}[h]
	\centering
	\includegraphics[width=\hsize]{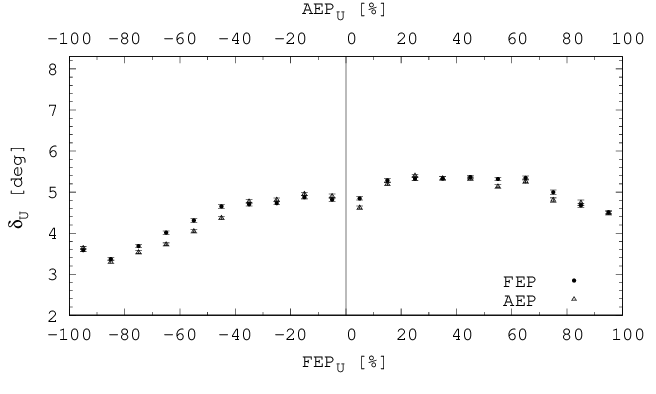}
	\caption{Spherical leading - following distance vs. FEP (measured on the bottom axis) plotted with black dots and AEP (measured on the top axis) depicted by empty triangles. The mean values of separation distance with their standard errors are averaged over 10 \% bins of the umbral FEP (AEP). The data of this figure comes from the SDD.}
	\label{Fig:gombitav}
\end{figure}

\noindent
Figure~\ref{Fig:gombitav} depicts the variation of the leading - following spherical distance as a function of FEP/AEP on the bottom/top axis, respectively. After the emergence the sunspot group starts expanding. This continues until the
AEP phase of -40 where the expansion reaches a plateau that extends to somewhat beyond the maximum area (AEP=0). After the peak area the separation distance increases again until about 60-70 \% of the decay. At the end of the decay the polarity separation shows a conspicuous decrease. These results are in agreement with the previous studies on the emergence and decay of ARs. \citet{2015LRSP...12....1V} summarized the process of the magnetic bipole emergence to the solar surface. When the tilted loop emerges by the buoyancy the footpoints are pushed apart, the legs of the $\Omega$ shape loop start to straighten which cause the increase of the separation distance between the two legs \citep{1995ApJ...441..886C}. The plateau is referred to as a stability plateau \citep{2003ApJ...593.1217P, 2023SSRv..219...64N} and \citet{2019A&A...625A..53S} reported that this can be observed 2.5 days after the beginning of the emergence. In the present study the beginning of the plateau is at -40 \% of AEP (30 \% of the entire lifetime), which corresponds to about 3 days in case of a lifetime of ten days. The expansion process has also been pointed out by \citet{2008ApJ...688L.115K, 2016A&A...596A...3V}.

\begin{figure}[h]
	\centering
	\includegraphics[width=\hsize]{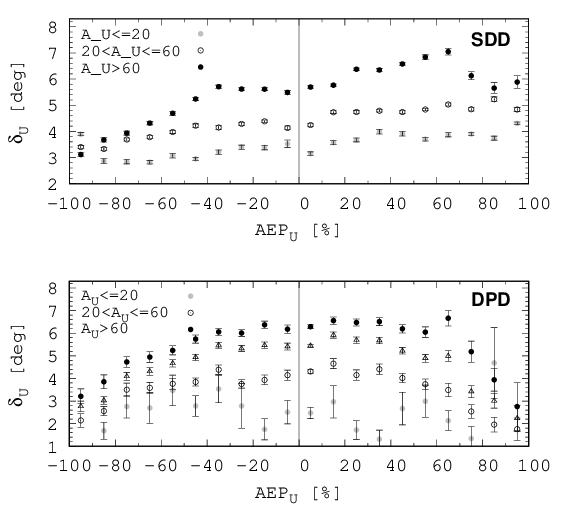}
	\caption{Separation distance as a function of AEP$_U$ for three sets of sunspot groups considering their peak umbral area. The top panel concerns the SDD, the bottom panel concerns the DPD. The empty triangles mark the data calculated on the entire sample of the DPD (same as in Figure~\ref{Fig:gombitav} for the SDD). The separation distances are averaged over 10\% bins of the AEP, the error bars mark the standard error.}
	\label{fig:gombitavter}
\end{figure}

\begin{figure}[h]
	\centering
	\includegraphics[width=\hsize]{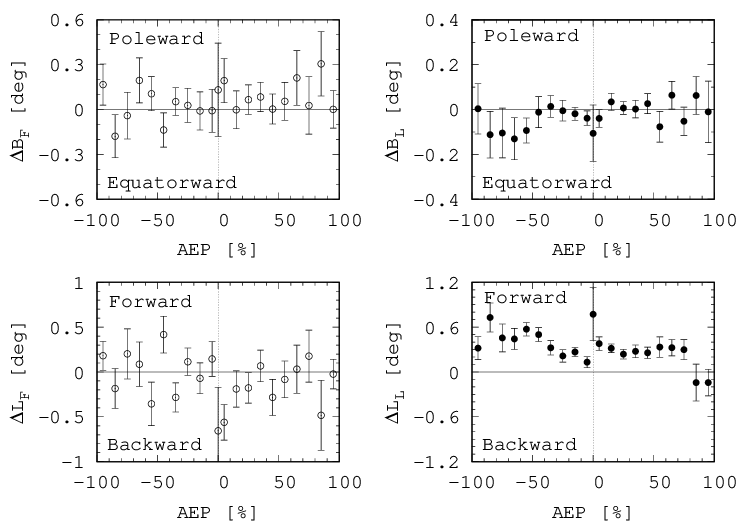}
	\caption{Daily latitudinal shift (top row) and longitudinal shift (bottom row) of the leading part (right column, plotted with dots) and following part (left column , depicted by circles) versus AEP. The shifts are averaged over each 10\% bin of AEP, the error bars mean the standard error. Positive/negative values of latitudinal shifts mean poleward/equatorward, and that of the longitudinal shifts concern forward/backward motion of sunspot groups.}
	\label{fig:shift}
\end{figure}

\noindent
Since the evolution of the active regions strongly depends on the peak area or peak flux, the variation of the separation distance is also analyzed by the area. Thus the studied groups are divided into three classes according to their maximum umbral areas (A$_U$): A$_U$ < 20 MSH, A$_U$ between 20 and 60 MSH and A$_U$ > 60 MSH (as in Figure~\ref{Fig:adpfdp}). Figure~\ref{fig:gombitavter} depicts these area divided data based on the SDD (top panel) and DPD (bottom panel). Sunspot groups of different areas evolve in different ways. The largest and middle size groups show the above mentioned path, while the leading - following separation of the smallest ones remains unchanged from the beginning of stable period and during the decay. The most pronounced decrease at the end of the decay phase is shown by the largest groups. The case is the same considering the DPD, although the plateaus in the growth phase are not as pronounced as in the case of the SDD, but the fallback in the end is much greater than that of the SDD's groups. The development of the smallest groups is much more chaotic and does not show increasing polarity separation, and their errors are higher than that of the larger groups or that of the SDD groups. This may be explained by the lack of magnetic data in the DPD where the leading - following parts are distinguished by the Kodaikanal method: these are the subgroups westward and eastward from the center of weight. However, \citet{2015MNRAS.447.1857B} showed that under a certain size the sunspot groups are unipolar and the leading - following distinction is irrelevant.
Since the leading - following expansion caused by the movement of the both parts their daily shift is also studied (Figure~\ref{fig:shift}). The most conspicuous difference between the upper and lower panels of Figure~\ref{fig:gombitavter} is that the distances measured after the 60\% decay time decrease more steeply in the data of DPD than in the SDD. This can be explained by the lack of polarity data in the DPD because when the trailing spots gradually disappear the program is increasingly restricted to the remaining spots in the leading part which is not the case with SDD data.

\noindent
The daily shift is determined as the difference between the latitudinal and longitudinal positions at two consecutive times extrapolated for 1 day. These shifts are averaged over 10 \% bins of AEP and calculated for the leading and following parts separately. The positive value of this shift concerns the forward and poleward motion. As one can see the the following part mainly moves backward (i.e., its values are in the negative half) and poleward (its values are in the positive half of the diagram) regardless the evolutionary phase of the group. This is not the case of the leading part. This part mainly show equatorward motion during the emergence phase but then remains at that latitudinal position. The longitudinal shift of the leading part (right bottom panel of Figure~\ref{fig:shift}) show the well-known forward motion during almost the entire life of the sunspot group, however at the end of its decay a pronounced backward motion can be observed. This means that the decrease in the separation distance at the end can be explained by the longitudinally backward motion and latitudinal immobility of the leading part. The biggest shift both in latitudinal and longitudinal directions is shown by the leading part at the beginning of the evolution. Similar findings have been reported by \citet{2019A&A...625A..53S} who showed that the emergence consists of two parts. The first is a fast movement that lasts until about 1 day after the emergence where the polarity separation accelerates and this is followed by a decelerating phase. The movements of the two opposite polarity parts are also opposite, i.e., the leading part moves forward and equatorward and the following part is shifted to backward and poleward directions.

\subsection{Long-Term Variation of the Separation Distance}
\begin{figure}[h]
	\centering
	\includegraphics[width=\hsize]{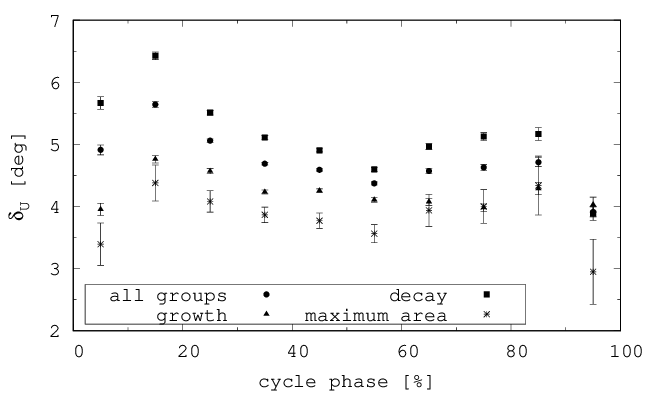}
	\caption{Separation distance versus cycle phase by using the SDD. The separation distances are averaged over 10 \% cycle phase bins and plotted to the center of each bin. The mean separation distance is calculated for the entire group (plotted by dots), their growth pi.e., the leading part moves forward and equatorward and the following part is phase (triangles), maxima (asterisk), decay (squares). The error bars mark the standard errors.}
	\label{fig:phasedep}
\end{figure}

The variation of the separation distance can also be studied on longer time interval than a half solar rotation. At first DPD and SDD data of cycle 23 have been used with similar results therefore only the results from the SDD are presented here.
The leading - following separation distance is calculated for all groups, and separately for the growth, maximum, and decay phases. Figure~\ref{fig:phasedep} shows these variations. The mean separation distance is highest during the decay phase and lowest at the maximum phase. The rate of the separation distance during the growth is between them. After a peak at the beginning of the cycle the polarity separation starts to decrease until about the half of the cycle. Then a slow increase can be observed. This variation also shows that the polarity separation strongly depends on the area and the larger the area of the groups the higher the separation distance between its two parts. The fallback around the middle of the cycle is caused by the so-called Gnevyshev-gap when the area of the solar ARs decreases temporarily. 

\begin{figure}[h]
	\centering
	\includegraphics[width=\hsize]{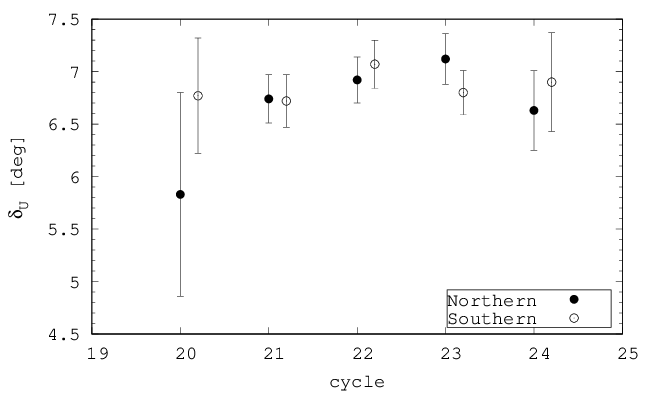}
	\includegraphics[width=\hsize]{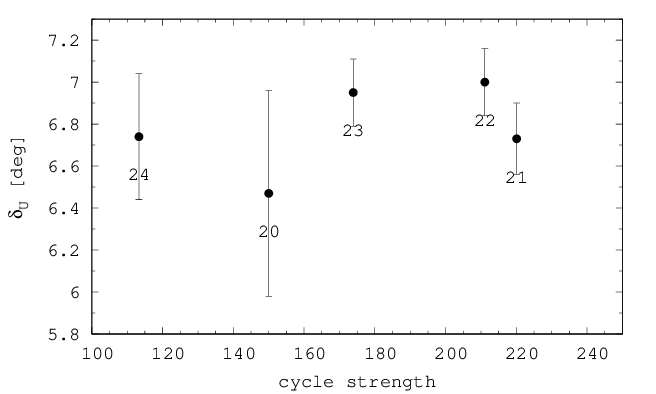}
	\caption{Hemispheric separation distances during SCs 20 - 24 (top panel). The separation distances are considered as their peak value and averaged over each solar cycle. Error bars represent the standard errors. Bottom panel: Maximum separation distances are averaged over solar cycles as a function of the cycle strengths. The latter is based on the yearly mean total sunspot number table created by WDC-SILSO, Royal Observatory of Belgium, Brussels. Error bars represent the standard errors, while the numbers mark the solar cycles.}
	\label{fig:cycle}
\end{figure}

\noindent
The top panel of Figure~\ref{fig:cycle} shows the hemispheric mean polarity separation averaged over the solar cycles by using the data from the DPD. The black dots mark the northern, while the empty circles depict the southern hemisphere's values. The most striking feature of this plot is the alternating cyclic variation of the hemispheric separation distance. During solar cycle 20 the large errors and the high distance between the northern and southern values is caused by the small number of data.

\noindent
Bottom panel of Figure~\ref{fig:cycle} depicts the cyclic mean separation distance as a function of the strength of the cycle. Each sunspot group is considered at the peak polarity separation distance's time, thus each group is considered only once.
The cycle strength data are come from the yearly mean total sunspot number table of the Royal Observatory of Belgium \footnote{\url{https://www.sidc.be/SILSO/datafiles\#total}}.
To the clear view it should be mention that the data of cycle 20 is not reliable because the starting date of DPD is 1974 while that of cycle 21 is 1976, thus there is a few data during cycle 20. The number of the studied distinct ARs is only 19 compared to that of the other cycles which is about 300. This data for solar cycle 24 is also low (89) and the standard error is relatively high, however the data of DPD covers almost the entire cycle 24. If we take all cycles into account the mean maximum separation distances does not show any correlation with the cycle strengths.

\section{Summary and conclusion}
The aim of this study was the investigation of the variation of leading - following separation distance on as large a sample as possible and checking its possible dependencies. The results are summarized below.
\begin{enumerate}
	\item The times of peak flux and peak area of the bulk of sunspot groups are the same, although there are some groups that have different peak flux and area times. In general the area of sunspot group reaches its peak value slightly earlier than its maximum flux. The growth phases last 2.56$\pm$0.02 days, 3.69$\pm$0.02 days and 4.96$\pm$0.03 days for the smallest, middle size and the largest groups, respectively. While the decay phases are 2.72$\pm$0.02 days, 4.13$\pm$0.02 days and 5.99$\pm$0.02 days for the smallest, middle size and the largest groups, respectively (Figure~\ref{Fig:adpfdp}).
	\item The separation distance of leading-following subgroups has a plateau beginning at the 30 \% of the sunspot group lifetime and after the maximum an abrupt jump causes a higher plateau. This behavior is independent of the lifetime of the groups and it is exhibited in both the area and magnetic flux development phases (Figure~\ref{Fig:gombitav}).
	\item At the end of the decay phase a shrinking can be observed, which mainly caused by the backward motion of the leading part. This decrease is more pronounced at the cases of the largest and middle sized groups and can not be observed on the smallest groups (Figures~\ref{fig:gombitavter} and \ref{fig:shift}). 
	\item The dependence of the separation distance on the area of ARs is strong, at the time of the Gnevyshev-gap the separation distance decreases temporarily (Figure~\ref{fig:phasedep}).
	\item The sign of difference between the hemispheric polarity separations alternate by the consecutive cycles (Figure~\ref{fig:cycle}).
\end{enumerate}

The emergence and decay of solar active regions are investigated in several works but the majority of these studies is restricted to short timescales and a few solar ARs \citep{ 2016A&A...596A...3V}. Some investigations used larger sample sizes, a few hundred sunspot groups \citep{2008ApJ...688L.115K, 2016ApJ...818....7M, 2019A&A...625A..53S}. Some of them, mainly the more recent studies focus on SDO measurements covering one cycle or one and a half cycles. The present uses the largest sample size on the longest time interval. Therefore not only the peak area or flux dependence of the separation distance can be studied, but also its cycle phase and cycle variations. 
\citet{2010ApJ...720..233C} using MHD simulation found that during the magnetic flux emergence the convective cells are elongated due to the horizontal expansion of the rising plasma and the parts of sunspot group with opposite polarities diverge. \citet{2012PhDT.......361H, 2015LRSP...12....1V} concluded that the peak separation follows the peak magnetic flux, and therefore the peak area, which is also corroborated by our results.   

\citet{2016ApJ...818....7M} assumed that larger sunspot groups are less influenced by the supergranular convection motion, while the smallest groups can not accumulate enough flux to prevail this disintegrating motion before their decay phase. Results in Figure~\ref{fig:gombitavter} shows the separation distances of the smallest groups begin to increase after their peak area, while those of the largest groups increase already during the emergence.
The poleward movement of the trailing parts observed mainly during the decay is in accordance with the flux transport dynamo theories. \citet{2010ApJ...720..233C} pointed out theoretically that the flux transport rate is inversely proportional to the magnetic field strength.
The reversal of the leading part at the end of the decay has also been found by \citet{2008ApJ...688L.115K} but it will be investigated further. \citet{2016A&A...596A...3V} described the expansion of one studied AR with an initial separation rate followed by a rapid expansion phase and a time span when the spread stalls and after that the AR slowly shrinks toward the end of the decay phase. Figures~\ref{Fig:gombitav} and \ref{fig:gombitavter} show these steps. At the beginning the speed of the separation is the fastest until AEP=-80\% which means about one day from the appearance of the group in the cases of the largest groups (bottom panel of Figure~\ref{Fig:adpfdp}). During this period, the rising $\Omega$ loop expands, the opposite polarity parts move apart. This phase is also described by \citet{2019A&A...625A..53S} who reported that this first phase lasts until about 1 day. Then the separation starts to slow down, this can be seen until about AEP=-40\% that corresponds to further two days in the largest groups. After the second phase the increase of separation is dragged until a temporary state of rest at the peak area. \citet{1995ApJ...441..886C} obtained theoretically that the emergence of an $\Omega$ loop to the surface is asymmetric, i.e., the leading leg of the flux rope is more inclined from the vertical direction. Thus we can suppose that the plateau can be observed when the leading leg becomes more vertical due to the buoyancy.
The next step is when the expansion is continuing. Here the disintegrating processes overcome the magnetic stability \citep{2016ApJ...818....7M}. The last phase of the sunspot group's lifetime is a converging motion caused by the eastward shift of the leading center (Figure~\ref{fig:shift}). 
\citet{1964suns.book.....B} showed this phenomenon near the maximum area of sunspot groups. Its physical background is still unclear as \citet{2015LRSP...12....1V} discusses based on the summary of \citet{2009LRSP....6....4F}.
According to \citet{2016ApJ...818....7M} the smallest groups can not accumulate enough magnetic flux to resist the supergranular convection thus their separation increases throughout their whole lifetimes in contrast to larger groups.\\                                                     No relationship can be pointed out between the cycle strength and separation distance on the studied sample (lower panel of Figure~\ref{fig:cycle}). This is also the case for the variation of the hemispheric separation distances if we consider the hemispheric strengths as the amplitudes of the corresponding cycle profiles \citep{2016ApJ...826..145M, 2020SoPh..295...86V}. If the cycle strength is defined as the amplitude of the centers of mass of each solar cycle (Figures 5 and 8 in \citet{2016ApJ...826..145M}) a slight relationship can be found between them. The amplitude of the northern hemispheric activity and the separation distance were slightly higher during SCs 21 and 23 while in SC 22 the case was opposite.
The possible nature of this hemispheric alternation of the leading - following separation distance is unclear, however this two Schwabe-cyclicity resembles to the $\alpha$ - $\Omega$ solar dynamo process with its 22-year duration just as assumed as the background of the Gnevyshev-Ohl rule \citep{2024ARep...68...89N}.

\citet{2008ssma.book.....S} showed that the lifetime of active regions is proportional to the magnetic flux and the rise time of the ARs decreases with the increasing area (\citet{2015LRSP...12....1V}). The rise time is found to be roughly 30 \% for ephemeral regions, and 15 \% for large regions containing sunspots and living longer than a week. In our results the emerging phases are 48 \%, 47 \% and 45 \% of the average lifetime of the studied groups from the smallest to the largest ones. Although the emergence phase of the sunspot groups is found to be shorter than the decay phase also in the present study but the relatively big discrepancy between the present results and those of \citet{2008ssma.book.....S} and \citet{2015LRSP...12....1V} might be explained by a methodological difference, I focused on umbrae while they consider the whole ARs, i.e., sunspot groups with pores, faculae and plages.

The above discussed topics will be studied on long-lived, recurrent sunspot groups and published in separate papers.

\begin{acknowledgements}
Thanks are due to the anonymous reviewer for her/his helpful advice. This research has received funding from National Research, Development and Innovation Office -- NKFIH, 141895. The cycle strength data source is WDC-SILSO, Royal Observatory of Belgium, Brussels.
\end{acknowledgements}

\bibliographystyle{aa} 
\bibliography{MURAKOZY_references4} 
\end{document}